\documentclass[aps,pra,twocolumn,showpacs,superscriptaddress]{revtex4-1}

\usepackage{amsfonts,mathrsfs,amsmath,amsthm,amssymb}
\usepackage{bm,times}
\usepackage{graphicx,epsfig}
\usepackage[colorlinks=true,breaklinks=true,linkcolor=blue,citecolor=blue,urlcolor=blue]{hyperref}

\hypersetup{
  pdfauthor={Tao Zhou},
  pdfkeywords={},
  pdftitle={Optimal programmable unambiguous discriminator between two unknown latitudinal states},
  pdfsubject={Research Paper},
  pdfpagemode=UseNone
}

\def \tr{ {\rm{Tr}}}
\def \non {\nonumber}

\def \ket {\rangle}
\def \bra {\langle}

\newcommand{\be}{\begin{eqnarray}}
\newcommand{\ee}{\end{eqnarray}}

\newcommand{\im}{\mathrm{i}}

\makeatletter
    
    \newcommand{\Rmnum}[1]{\expandafter\@slowromancap\romannumeral #1@}
\makeatother

\begin{document}
\title{Optimal programmable unambiguous discriminator between two unknown latitudinal states}
\author{Xiaobing Sunian}
\affiliation{School of Physical Science and Technology, Southwest Jiaotong University, Chengdu 610031, China}
\author{Yuwei Li}
\affiliation{School of Physical Science and Technology, Southwest Jiaotong University, Chengdu 610031, China}
\author{Tao Zhou}
\email{taozhou@swjtu.edu.cn}
\affiliation{School of Physical Science and Technology, Southwest Jiaotong University, Chengdu 610031, China}

\date{\today}

\begin{abstract}
Two unknown states can be unambiguously distinguished by a universal programmable discriminator, which has been widely discussed in previous works and the optimal solution has also been obtained. In this paper, we investigate the programmable unambiguous discriminator between two unknown ``latitudinal'' states, which lie in a subspace of the total state space. By equivalence of unknown pure states to known average mixed states, the optimal solution for this problem is systematically derived, and the analytical success probabilities for the optimal unambiguous discrimination are obtained. It is beyond one's expectation that the optimal setting for the programmable unambiguous discrimination between two unknown ``latitudinal'' states is the same as that for the universal ones. The results in this work can be used for the realization of the programmable discriminator in laboratory.
\end{abstract}


\maketitle

\section{Introduction}
Quantum state discrimination is required for many applications in quantum communication and quantum computation, and therefore plays an essential role in quantum information theory, \text{i. e.} quantum key distribution~\cite{BB84,PhysRevA.65.032302,PhysRevA.68.042315,PhysRevA.78.022321}, quantum secret sharing~\cite{PhysRevA.59.1829}, quantum scure direct communication~\cite{PhysRevA.68.042317,PhysRevA.69.052319,PhysRevA.71.044305,Gu2011,Chang2013,Zheng2014,Zou2014}, and other quantum information processing~\cite{PhysRevA.93.020101,Ai2016,PhysRevA.92.042314,Xu2015,Dong2015,Long2014}. The discrimination between quantum states is not a trivial problem in quantum mechanics since an unknown state can not be cloned perfectly~\cite{Nat10.5886,0256-307X-28-1-010306}. Two basic strategies have been introduced to achieve quantum state discrimination: one is the minimum-error discrimination (MD)~\cite{Quan_Helstrom,Pro_Holevo,PhysRevA.64.030303,PhysRevA.65.052308,PhysRevA.68.042305,PhysRevA.65.050305}, with a minimal probability for the error, and the other one is the unambiguous discrimination (UD)~\cite{IVANOVIC1987257,DIEKS1988303,PERES198819,Jaeger199583,Chefles1998339,PhysRevA.79.052302}, with a minimum probability of inconclusive results. In those works, a quantum state is chosen from a set of states to be identified, and one do not know which, and wants to determine the actual states. The set of states to be distinguished is known, and the device for the state discrimination is not universal but specifically designed for the states to be distinguished. 

On the other hand, illuminated by the spirit of programmable quantum devices~\cite{PhysRevLett.79.321,PhysRevLett.89.190401,PhysRevLett.94.090401,PhysRevA.65.022301,PhysRevA.66.022112,PhysRevA.69.032302,PhysRevA.69.052321}, it is meaningful to construct a \textit{universal} quantum device that can unambiguously discriminate between quantum states, which means, the setting of the device is not dependent on the input states compared with the usual state discrimination before. For qubit case, such a universal discriminator was first constructed by Bergou and Hillery~\cite{PhysRevLett.94.160501}, and in this programmable device, there are two program registers denoted by $A$ and $C$, respectively, and a data register denoted by $B$. It is assumed that the program registers $A$ and $C$ are prepared in the qubit states $|\psi_1\ket$ and $|\psi_2\ket$, respectively, while the data register $B$ is prepared in a third state (guaranteed to be either $|\psi_1\ket$ or $|\psi_2\ket$ with \textit{a priori} probabilities $\eta_1$ and $\eta_2$ such that $\eta_1+\eta_2=1$) that one would like to identify. This universal discriminator can measure the total input states
\begin{subequations}
\label{instates}
\be
|\Psi_1\ket&=&|\psi_1\ket_A|\psi_1\ket_B|\psi_2\ket_C\\
|\Psi_2\ket&=&|\psi_1\ket_A|\psi_2\ket_B|\psi_2\ket_C,
\ee
\end{subequations}
and it is a amazing feature of this device that the states in the registers could be completely unknown compared with the usual state discrimination. This discriminator can distinguish any pair of states in this device with some probability, unless the two states $|\psi_1\ket$ and $|\psi_2\ket$ are identical.

The generalizations and the experimental realizations of this programmable discriminator above were also introduced and widely discussed soon~\cite{PhysRevA.73.062334,Zhou2011,QIC12.1017,PhysRevA.73.012328,PhysRevA.82.042312,*PhysRevA.83.039909,PhysRevA.72.052306,PhysRevA.74.042308,PhysRevA.75.032316,PhysRevA.75.052116,PhysRevA.76.032301,PhysRevA.77.034306,PhysRevA.78.032320,*PhysRevA.78.069902,PhysRevA.78.042315,PhysRevA.80.042311,PhysRevA.83.052328,He2006103,PhysRevA.88.052304,PhysRevA.89.014301}. In these works, either the minimum-error strategy or the unambiguous discrimination strategy was considered, and the two strategies were unified by introducing an error margin~\cite{PhysRevA.88.052304}. The cases for the multiple copies of input states in both program and data registers are discussed, and the optimal solutions are obtained for $n_A=n_C=n, n_B=1$~\cite{PhysRevA.73.012328,He2006103}, for $n_A=n_C=1,n_B=n$~\cite{PhysRevA.73.062334}, for $n_A=n_C=n,n_B=m$~\cite{PhysRevA.75.032316}, and for arbitrary copies of states in both program and data registers~\cite{PhysRevA.82.042312,PhysRevA.83.052328,PhysRevA.89.014301}, where $n_A$, $n_B$ and $n_C$ are the copies of states in the registers $A$, $B$ and $C$, respectively. The cases for high dimensional (qudit) states in the registers were also considered, and the unambiguous discrimination has been discussed for $n_A=n_B=n_C=1$~\cite{Zhou2011}. Furthermore, the case where each copy in the registers is a mixed state was also treated~\cite{PhysRevA.82.042312}. The most general case for the pure state is the copies of each registers are all arbitrary, and moreover, the states in the registers are qudit states rather than qubit states only, say, the task is to distinguish the two total input states
\begin{subequations}
\be
|\Psi_1\ket&=&|\psi_1\ket_A^{\otimes n_A}|\psi_1\ket_B^{\otimes n_B}|\psi_2\ket_C^{\otimes n_C}\\
|\Psi_2\ket&=&|\psi_1\ket_A^{\otimes n_A}|\psi_2\ket_B^{\otimes n_B}|\psi_2\ket_C^{\otimes n_C},
\ee
\end{subequations}
where $|\psi_1\ket$ and $|\psi_2\ket$ are two unknown pure states in $d$-dimensional $(d\geqslant2)$ Hilbert space. For this case, the optimal solutions for both the minimum-error discrimination and unambiguous discrimination were obtained in a group-theoretic approach~\cite{QIC12.1017} and the success probabilities are systematically derived for the unambiguous discrimination~\cite{PhysRevA.89.014301}, while only the minimum-error discrimination was discussed in Ref.~\cite{PhysRevA.83.052328}.

In the present paper, we investigate a special version of the programmable unambiguous discrimination between two qubit states, where the unknown input states are restricted on the latitude line of the Bloch sphere, not the whole state space, say
\be
\label{latistate}
|\psi\ket=\cos(\theta/2)|0\ket+e^{-\im\phi}\sin(\theta/2)|1\ket,
\ee
with $\theta$ a fixed number in $[0,\pi]$ and $\phi$ uniformly distributed in $[0,2\pi)$, and $\{|0\ket,|1\ket\}$ represent a basis for the space $\mathcal{H}$ of qubit states. We call states of this form the \textit{latitudinal} states, and for a specific case $\theta=\pi/2$, one has the ``equatorial'' states~\cite{PhysRevA.62.012302}
\be
|\psi\ket=\frac{1}{\sqrt{2}}(|0\ket+e^{-\im\phi}|1\ket).
\ee
Studying the programmable unambiguous discrimination for the restricted states to the latitude is quite reasonable, and this is motivated by the physical implementations of quantum communication as well as by the fundamental questions in quantum information processing. For instance,   all existing quantum cryptographic experiments are using states that are on the equator, rather than states that span the whole Bloch sphere. Comparing with the universal programmable unambiguous discrimination, there is more information about the states since they are restricted to a latitude, and one may take it for granted that the optimal solution is different for the programmable unambiguous discrimination between two unknown ``latitudinal'' states, and conjecture that higher success probabilities could be obtained for the equatorial states. To verify the validity of these conjectures is the main purpose of this work.

This paper is organized as follows. In Sec.~\ref{Sec2}, the equivalent average mixed states of the input states are derived for the programmable unambiguous discrimination, and he structure of the average input mixed states is briefly discussed. In Sec.~\ref{Sec3}, the optimal setting is obtained via the unambiguous discriminations between mixed states in the subspaces, and the optimal success probabilities for the average input mixed states are given. The success probabilities for the optimal programmable unambiguous discriminator between two unknown latitudinal states are given in Sec.~\ref{Sec4}. We end this paper with a short discussion in Sec.~\ref{Sec5}.

\section{Average input states as equivalent mixed states}
\label{Sec2}
In this section, we will discuss the equivalence of the unknown pure states to the known mixed states. For the two input states
\begin{subequations}
\be
|\psi_1\ket&=&\cos(\theta/2)|0\ket+e^{-\im\phi_1}\sin(\theta/2)|1\ket\\
|\psi_2\ket&=&\cos(\theta/2)|0\ket+e^{-\im\phi_2}\sin(\theta/2)|1\ket,
\ee
\end{subequations}
with $\phi_1$ and $\phi_2$ uniformly distributed in $[0,2\pi)$, though restricted to a latitude line of the Bloch sphere, they are still unknown to us, and they can change by different preparations. The permutation symmetry properties of the total input states $|\Psi_1\ket$ and $|\Psi_2\ket$ is preserved and this is the available information to distinguish $|\Psi_1\ket$ and $|\Psi_2\ket$. Therefore, one can introduce two average mixed states for the pure input states in Eqs.~(\ref{instates}),
\begin{subequations}
\label{inmixed}
\be
\rho_1&=&\frac{1}{(2\pi)^2}\int_0^{2\pi}\int_0^{2\pi}|\Psi_1\ket\bra\Psi_1|d\phi_1d\phi_2\\
\rho_2&=&\frac{1}{(2\pi)^2}\int_0^{2\pi}\int_0^{2\pi}|\Psi_2\ket\bra\Psi_2|d\phi_1d\phi_2.
\ee
\end{subequations}
It is obvious that the optimal strategy for discrimination between the two pure states in Eq.~(\ref{instates}) is also optimal on average, and the discrimination of unknown pure states is, on average, equivalent to the discrimination between the known average mixed states in Eq.~(\ref{inmixed}). We can unambiguously discriminate between $|\Psi_1\ket$ and $|\Psi_2\ket$ as soon as we can unambiguously discriminate between $\rho_1$ and $\rho_2$, and before this, more explicit expressions for the average states $\rho_1$ and $\rho_2$ should be given first. 

For the state $|\psi\ket$ in Eq.~(\ref{latistate}), it is easy to obtain the following two facts
\be
\frac{1}{2\pi}\int_0^{2\pi}|\psi\ket\bra\psi|d\phi=\cos^2\frac{\theta}{2}|0\ket\bra0|+\sin^2\frac{\theta}{2}|1\ket\bra1|,
\ee
and 
\be
\frac{1}{2\pi}\int_0^{2\pi}|\psi\psi\ket\bra\psi\psi|d\phi&=&\cos^4\frac{\theta}{2}|00\ket\bra00|+\sin^4\frac{\theta}{2}|11\ket\bra11|\non\\
&&+2\cos^2\frac{\theta}{2}\sin^2\frac{\theta}{2}|u\ket\bra u|,
\ee
where
\be
|u\ket=\frac{1}{\sqrt{2}}\big(|01\ket+|10\ket\big),
\ee
and one has
\begin{subequations}
\label{avstates}
\be
\rho_1&=&\cos^6\frac{\theta}{2}|000\ket\bra000|+\cos^4\frac{\theta}{2}\sin^2\frac{\theta}{2}|001\ket\bra001|\non\\
&&+\cos^2\frac{\theta}{2}\sin^4\frac{\theta}{2}|110\ket\bra110|+\sin^6\frac{\theta}{2}|111\ket\bra111|\non\\
&&+2\cos^4\frac{\theta}{2}\sin^2\frac{\theta}{2}|u\ket\bra u|\otimes|0\ket\bra0|\non\\
&&+2\cos^2\frac{\theta}{2}\sin^4\frac{\theta}{2}|u\ket\bra u|\otimes|1\ket\bra1|,\\
\rho_2&=&\cos^6\frac{\theta}{2}|000\ket\bra000|+\cos^4\frac{\theta}{2}\sin^2\frac{\theta}{2}|100\ket\bra100|\non\\
&&+\cos^2\frac{\theta}{2}\sin^4\frac{\theta}{2}|011\ket\bra011|+\sin^6\frac{\theta}{2}|111\ket\bra111|\non\\
&&+2\cos^4\frac{\theta}{2}\sin^2\frac{\theta}{2}|0\ket\bra 0|\otimes|u\ket\bra u|\non\\
&&+2\cos^2\frac{\theta}{2}\sin^4\frac{\theta}{2}|1\ket\bra 1|\otimes|u\ket\bra u|.
\ee
\end{subequations}

Let $H_1$ be the Hilbert space of $\rho_1$, which is spanned by $\{|000\ket,|001\ket,|u\ket|0\ket,|u\ket|1\ket,|110\ket,|111\ket\}$, and $H_2$ the Hilbert space of $\rho_2$, which is spanned by $\{|000\ket,|100\ket,|0\ket|u\ket,|1\ket|u\ket,|011\ket,|111\ket\}$. Define $\mathcal{H}_0$ the space spanned by $\{|000\ket,|111\ket\}$, $\mathcal{H}_1$ the space spanned by $\{|001\ket,|u\ket|0\ket,|100\ket,|0\ket|u\ket\}$, and $\mathcal{H}_2$ the space spanned by $\{|110\ket,|u\ket|1\ket,|011\ket,|1\ket|u\ket\}$. One should notice that both $\mathcal{H_1}$ and $\mathcal{H}_2$ are 2-dimensional subspaces. It is obvious the total space $H=\mathcal{H}^{\otimes 3}=H_1\cup H_2=\mathcal{H}_0\oplus\mathcal{H}_1\oplus\mathcal{H}_2$, and therefore, the discrimination between $\rho_1$ and $\rho_2$ reduces to the state discriminations in each subspaces $\mathcal{H}_0$,  $\mathcal{H}_1$, and $\mathcal{H}_2$. The details for the discriminations in the subspaces are discussed in the following section.

\section{Optimal unambiguous discriminations in subspaces}
\label{Sec3}
From Eq.~(\ref{avstates}), the average mixed states $\rho_1$ and $\rho_2$ are the same in the subspace $\mathcal{H}_0$, and thus can not be distinguished further in this subspace. In the subspace $\mathcal{H}_1$, the mixed state
\be
\rho'_1=\frac{1}{3}|001\ket\bra001|+\frac{2}{3}|u\ket\bra u|\otimes|0\ket\bra0|.
\ee
occurs with probability $3\cos^4\frac{\theta}{2}\sin^2\frac{\theta}{2}$ for $\rho_1$, while  for $\rho_2$, the mixed state
\be
\rho'_2=\frac{1}{3}|100\ket\bra100|+\frac{2}{3}|0\ket\bra0|\otimes|u\ket\bra u|.
\ee
occurs with the same probability $3\cos^4\frac{\theta}{2}\sin^2\frac{\theta}{2}$. Therefore, for the discrimination, the probability for the occurrence of a vector in space $\mathcal{H}_1$ is 
\be
p_1=3\cos^4\frac{\theta}{2}\sin^2\frac{\theta}{2},
\ee
and finally, in subspace $\mathcal{H}_1$, the problem reduces to the unambiguous discrimination between two mixed states $\rho'_1$ and $\rho'_2$ occurring with probabilities $\eta_1$ and $\eta_2$, respectively. The measurement procedure for the unambiguous discrimination between $\rho'_1$ and $\rho'_2$ has three outcomes, associated with identifying the state as $\rho'_1$, identifying the state as $\rho'_2$, and failing to identify the state, and is mathematically represented by three POVM elements $E_1^{}$, $E_2$ and $E_0=\mathbb{I}-E_1-E_2$, where $\mathbb{I}$ is the identity operator on the space $\mathcal{H}_1$. It is required that no error can happen in the discrimination, so
\be
\label{cond1}
\tr(\rho'_1E_2)=\tr(\rho'_2E_1)=0.
\ee
The success probability of the unambiguous discrimination is 
\be
\label{sucp}
P_1=\eta_1\tr(\rho'_1E_1)+\eta_2\tr(\rho'_2E_2).
\ee
The optimal solution is to maximize Eq.~(\ref{sucp}) subject to Eq.~(\ref{cond1}) and the constraint that $E_0$, $E_1$, and $E_2$ are semi-positive definite.

Before the optimal POVM can be constructed, one need to further consider the structure of $\mathcal{H}_1$. Define $\mathbb{H}_1$ the space spanned by $\{|001\ket,|u\ket|0\ket\}$, and $\mathbb{H}_2$ the space spanned by $\{|100\ket,|0\ket|u\ket\}$, and then, $\mathbb{H}_1=\sup\rho'_1$, $\mathbb{H}_2=\sup\rho'_2$, $\mathcal{H}_1=\mathbb{H}_1\bigcup\mathbb{H}_2$. The unambiguity conditions in Eq.~(\ref{cond1}) means that the support of $E_2$ ($E_1$) is a subspace of the kernel of $\rho'_1$ ($\rho'_2$)~\cite{PhysRevA.68.010301,PhysRevA.71.050301,PhysRevA.81.020304}, and then $\sup E_1\subset\ker\rho'_2$ and $\sup E_2\subset\ker\rho'_1$. Due to the structure of $\mathcal{H}_1$, $\ker\rho'_1$ is a one-dimensional subspace spanned by $|v\ket|0\ket$, and $\ker\rho'_2$ is a one-dimensional subspace spanned by $|0\ket|v\ket$, where
\be
|v\ket=\frac{1}{\sqrt{2}}\big(|01\ket-|10\ket\big).
\ee
The POVM elements $E_1$ and $E_2$ now are
\be
E_1=\alpha|0\ket\bra0|\otimes|v\ket\bra v|,\ \ E_2=\beta|v\ket\bra v|\otimes|0\ket\bra0|,
\ee
with $0\leqslant\alpha,\beta\leqslant1$ guaranteeing the semi-positive definite properties. In the orthonormal basis $\{|001\ket,|u\ket|0\ket,|v\ket|0\ket\}$ of $\mathcal{H}_1$, the POVM element $E_0$ is given by the $3\times3$ matrix
\be
E_0=\left(\begin{array}{ccc}
1-\alpha/2 & \alpha/2\sqrt{2} & \alpha/2\sqrt{2}\\
\alpha/2\sqrt{2} & 1-\alpha/4  & -\alpha/4\\
\alpha/2\sqrt{2} & -\alpha/4& 1-\alpha/4-\beta
\end{array}\right),
\ee
and its eigenvalues are easy to obtain explicitly, and required to be nonnegativity due to the semi-positive definite condition. It can be directly derived from the matrix above,
\be
\beta\leqslant\frac{4-4\alpha}{4-3\alpha}.
\ee
For the success probability in Eq.~(\ref{sucp}), 
\be
P_1&=&\frac{1}{3}(\eta_1\alpha+\eta_2\beta)\non\\
&\leqslant&\frac{1}{3}(\eta_1\alpha+\eta_2\frac{4-4\alpha}{4-3\alpha})\non\\
&\leqslant&\frac{4}{9}(1-\sqrt{\eta_1\eta_2}),
\ee
where the equalities hold for 
\be
\alpha=\frac{2}{3}\bigg(2-\sqrt{\frac{\eta_2}{\eta_1}}\bigg),\ \ \beta=\frac{2}{3}\bigg(2-\sqrt{\frac{\eta_1}{\eta_2}}\bigg).
\ee 
The constrains $0\leqslant\alpha,\beta\leqslant1$ inform us the results above work only for
\be
\frac{1}{5}\leqslant\eta_1\leqslant\frac{4}{5}.
\ee
For $0\leqslant\eta_1<1/5$, the optimal solution is 
\be
P_1^{\rm opt}=\frac{1}{3}\eta_2
\ee
with $\alpha=0,\beta=1$,
and for $4/5<\eta_1\leqslant1$, the optimal solution is 
\be
P_1^{\rm opt}=\frac{1}{3}\eta_1,
\ee
with $\alpha=1,\beta=0$.
The results for the unambiguous discrimination in space $\mathcal{H}_1$ can be summarized as follow
\be
P_1^{\rm opt}=\left\{\begin{array}{ll}
\frac{\displaystyle1}{\displaystyle 3}\eta_2 & \eta_1<\frac{\displaystyle
1}{\displaystyle 5}\\ \\
\frac{\displaystyle 4}{\displaystyle 9}(1-\sqrt{\eta_1\eta_2}) & \frac{\displaystyle 1}{\displaystyle 5}\leqslant \eta_1\leqslant\frac{\displaystyle 4}{\displaystyle 5}\\ \\
\frac{\displaystyle 1}{\displaystyle 3}\eta_1 & \eta_1>\frac{\displaystyle 4}{\displaystyle5}
\end{array} \right..
\ee

For the subspace $\mathcal{H}_2$, the mixed state
\be
\rho''_1=\frac{1}{3}|110\ket\bra110|+\frac{2}{3}|u\ket\bra u|\otimes|1\ket\bra1|.
\ee
occurs with probability $3\cos^2\frac{\theta}{2}\sin^4\frac{\theta}{2}$ for $\rho_1$, while  for $\rho_2$, the mixed state
\be
\rho''_2=\frac{1}{3}|011\ket\bra011|+\frac{2}{3}|1\ket\bra1|\otimes|u\ket\bra u|.
\ee
occurs with the same probability $3\cos^2\frac{\theta}{2}\sin^4\frac{\theta}{2}$. Therefore,  the probability for the occurrence of a vector in space $\mathcal{H}_2$ is 
\be
p_2=3\cos^2\frac{\theta}{2}\sin^4\frac{\theta}{2},
\ee
Similar discussions can be carried on for the unambiguous discrimination between $\rho''_1$ and $\rho''_2$ in the subspace $\mathcal{H}_2$, and the POVM elements have the form
\be
E'_1&=&\alpha'|1\ket\bra1|\otimes|v\ket\bra v|,\ \ E'_2=\beta'|v\ket\bra v|\otimes|1\ket\bra1|,\non\\
E'_0&=&\mathbb{I}'-E'_1-E'_2,
\ee
where $\mathbb{I}'$ is the identity operator on $\mathcal{H}_2$. The optimal success probability is
\be
P_2^{\rm opt}=\frac{4}{9}(1-\sqrt{\eta_1\eta_2})
\ee
for $1/5\leqslant\eta_1\leqslant4/5$ with $\alpha'=\frac{2}{3}\big(2-\sqrt{\frac{\eta_2}{\eta_1}}\big), \beta'=\frac{2}{3}\big(2-\sqrt{\frac{\eta_1}{\eta_2}}\big)$. For $0\leqslant\eta_1<1/5$,
\be
P_2^{\rm opt}=\frac{1}{3}\eta_2
\ee
with $\alpha'=0,\beta'=1$ and for $4/5<\eta_1\leqslant1$
\be
P_2^{\rm opt}=\frac{1}{3}\eta_1,
\ee
with $\alpha'=1,\beta'=0$.

Following the results for the unambiguous discrimination in both subspaces $\mathcal{H}_1$ and $\mathcal{H}_2$, the POVM for the unambiguous discrimination between $\rho_1$ and $\rho_2$ is 
\be
\Pi_1&=&E_1+E'_1,\non\\
\Pi_2&=&E_2+E'_2,\non\\
\Pi_0&=&\openone-\Pi_1-\Pi_2,
\ee
where $\openone$ is the identity operator on the total space $H$. The optimal solution for the POVM is 
\be
\label{optpovm}
\Pi_1=c_1\mathbf{I}\otimes|v\ket\bra v|,\ \ \Pi_2=c_2|v\ket\bra v|\otimes \mathbf{I}
\ee
with $\mathbf{I}$ the identity operator on the space $\mathcal{H}$, where for $1/5\leqslant\eta_1\leqslant4/5$, $c_1=\frac{2}{3}\big(2-\sqrt{\frac{\eta_2}{\eta_1}}\big), c_2=\frac{2}{3}\big(2-\sqrt{\frac{\eta_1}{\eta_2}}\big)$; for $0\leqslant\eta_1<1/5$, $c_1=0,c_2=1$; and for $4/5<\eta_1\leqslant1$, $c_1=1,c_2=0$. Finally, the optimal success probability for the unambiguous discrimination between the average states $\rho_1$ and $\rho_2$ is
\be
P^{\rm opt}&=&p_1P_1^{\rm opt}+p_2P_2^{\rm opt}\non\\
&=&\left\{\begin{array}{ll}
\eta_2\cos^2\frac{\theta}{2}\sin^2\frac{\theta}{2} & \eta_1<\frac{\displaystyle
1}{\displaystyle 5}\\ \\
\frac{\displaystyle 4}{\displaystyle 3}\cos^2\frac{\theta}{2}\sin^2\frac{\theta}{2}(1-\sqrt{\eta_1\eta_2}) & \frac{\displaystyle 1}{\displaystyle 5}\leqslant \eta_1\leqslant\frac{\displaystyle 4}{\displaystyle 5}\\ \\
\eta_1\cos^2\frac{\theta}{2}\sin^2\frac{\theta}{2} & \eta_1>\frac{\displaystyle 4}{\displaystyle5}
\end{array} \right..
\ee

\section{Success probabilities for the optimal programmable discriminator}
\label{Sec4}
The optimal setting for the unambiguous discriminator between two unknown latitudinal states has been given in Eq.~(\ref{optpovm}), and one can conclude the optimal measurement is a POVM for $1/5\leqslant\eta_1\leqslant4/5$, and a projector measurement for both $\eta_1<1/5$ and $\eta_1>4/5$. Actually, the optimal setting here is exactly the same as that for the universal ones, which is beyond our expectation. In fact, the unknown states rather than the average mixed states are distinguished in the device, and therefore we will derive the success probability for the pure input states in this section.

With the expressions of the optimal POVM operators in Eq.~(\ref{optpovm}), one can have
\be
\bra\Psi_1|\Pi_1|\Psi_1\ket&=&c_1\big|\bra\psi_1|\bra\psi_2|v\ket\big|^2\non\\
&=&c_1\cos^2\frac{\theta}{2}\sin^2\frac{\theta}{2}\big[1-\cos(\phi_1-\phi_2)\big].
\ee
Meanwhile,
\be
\big|\bra\psi_1|\psi_2\ket\big|^2&=&\cos^4\frac{\theta}{2}+\sin^4\frac{\theta}{2}+2\cos^2\frac{\theta}{2}\sin^2\frac{\theta}{2}\cos(\phi_1-\phi_2)\non\\
&=&1-2\cos^2\frac{\theta}{2}\sin^2\frac{\theta}{2}\big[1-\cos(\phi_1-\phi_2)\big],
\ee
and therefore
\be
\bra\Psi_1|\Pi_1|\Psi_1\ket=\frac{c_1}{2}\big(1-\big|\bra\psi_1|\psi_2\ket\big|^2\big).
\ee
Similarly, one can have
\be
\bra\Psi_2|\Pi_2|\Psi_2\ket=\frac{c_2}{2}\big(1-\big|\bra\psi_1|\psi_2\ket\big|^2\big).
\ee
Finally, we can come to the optimal success probability for the pure input states
\be
P_{\rm suc}^{\rm opt}&=&\eta_1\bra\Psi_1|\Pi_1|\Psi_1\ket+\eta_2\bra\Psi_2|\Pi_2|\Psi_2\ket\non\\
&=&\left\{\begin{array}{ll}
\frac{\displaystyle 1}{\displaystyle 2}\eta_2\big(1-\big|\bra\psi_1|\psi_2\ket\big|^2\big) & \eta_1<\frac{\displaystyle
1}{\displaystyle 5}\\ \\
\frac{\displaystyle 2}{\displaystyle 3}(1-\sqrt{\eta_1\eta_2})\big(1-\big|\bra\psi_1|\psi_2\ket\big|^2\big) & \frac{\displaystyle 1}{\displaystyle 5}\leqslant \eta_1\leqslant\frac{\displaystyle 4}{\displaystyle 5}\\ \\
\frac{\displaystyle 1}{\displaystyle 2}\eta_1\big(1-\big|\bra\psi_1|\psi_2\ket\big|^2\big) & \eta_1>\frac{\displaystyle 4}{\displaystyle5}
\end{array} \right.,\non\\
\ee
and this expression is also the same as that in Ref.~\cite{PhysRevLett.94.160501}.

\section{Conclusions and discussions}
\label{Sec5}

In conclusion, we have investigated the programmable unambiguous discriminator between two unknown latitudinal states. The equatorial states are a special class of the latitudinal states, which has been widely used in quantum communications theoretically and experimentally. By the equivalence of unknown pure states to the mean mixed states, the discrimination problem reduces to the unambiguous discriminations between known two mixed states in subspaces. The Jordan-basis approach does not work here, and scheme for unambiguous discrimination between  mixed states are applied in the subspaces. Actually, it is more complicated to design a programmable device for the latitudinal states than the universal ones, since there are more symmetry properties for the average mixed states for the universal programmable discrimination. The optimal setting is systematically derived, and the optimal success probabilities are obtained, which are the same as that for the universal ones in previous work~\cite{PhysRevLett.94.160501}. The expressions of the optimal measurement operators are given, and this is very useful and helpful in the construction for the programmable discriminator.  The results in this work suggest that only the permutation properties of the input states are useful to design the programmable discriminator between the unknown states, and it is interesting to provide a strict demonstration for this conclusion elsewhere. We expect that our results could come up with further theoretical or experimental consequences.

\acknowledgements
This work was supported by the National Natural Science Foundation of China under Grants No.~11405136 and No.~11547311, and the Fundamental Research Funds for the Central Universities under Grants No.~2682016CX059 and No.~2682014BR056.

\bibliography{refs}

\end{document}